\documentclass[
aps,%
12pt,%
final,%
notitlepage,%
oneside,%
onecolumn,%
nobibnotes,%
nofootinbib,%
superscriptaddress,%
noshowpacs,%
centertags]%
{revtex4}

\begin{document}

\newcounter{myfigure}

\title {THE BRIGHTEST OH MASER IN THE SKY: A FLARE OF EMISSION IN  W75~N.}

\author{\firstname{A.V.} \surname{Alakoz}}
\email{rett@tanatos.asc.rssi.ru}

\author{\firstname{V.I.} \surname{Slysh}}

\author{\firstname{M.V.} \surname{Popov}}

\author{\firstname{I.E.} \surname{Val'tts}}

\affiliation{%
Astro Space Center of the Lebedev Physical Center of RAS,
Profsoyuznaya Str. 84/32, 117997 Moscow, Russia
}%

\begin{abstract}

A flare of maser radio emission in the OH-line 1665~MHz has been discovered in the star
forming region W75~N in 2003, with the flux density of about 1000~Jy. At the time
it was the strongest OH maser detected during the whole history of observations since the discovery
of cosmic masers in 1965.
The flare emission is linearly polarized with a degree of polarization near 100\%.
A weaker flare with a flux of 145~Jy was observed in this source in 2000~--~2001,
which was probably a precursor of the powerful flare.   Intensity
of two other spectral features has
decreased after beginning of the flare. Such variation of the intensity of maser
condensation emission (increasing of one and decreasing of the other)
can be explained by passing of the magneto hydrodynamic
shock across  regions of  enhanced gas concentration.

\end{abstract}

\maketitle

\noindent


\clearpage
\section*{Introduction}

W75~N is one of the active star formation regions in the Galaxy
which is accompanied by the maser emission in lines of OH, methanol and H$_2$O.
Mapping observations of the masers with high angular resolution has shown that
the maser emission sources are connected with ultracompact HII regions
(\citealp{hunter};~\citealp{minier};~\citealp{slysh02}), and
possibly are located in protoplanetary disk surrounding massive stars.
In this paper discovery of the powerful flare of emission is reported
by which W75~N became the brightest OH cosmic maser for the whole history of observations.

\section*{Observations}
Polarization observations of the maser emission from W75~N in all
four Stokes parameters were conducted at the Nan\c{c}ay radio telescope (France)
(\citealp{driel}) in October~2003, and in two circular polarizations at the
64-m radio telescope in Kalyazin (Russia) (\citealp{kalyazin}) in July and October~2004,
in OH lines 1665 and 1667~MHz. Digital autocorrelators with spectral resolution 0.137~km/s
were used at both telescopes. The number of channels at Nan\c{c}ay was 8192, divided in
8 sections by 1024~channels each (both frequencies were observed, and each section corresponded
to a separate polarization), the bandwidth was 0.781~MHz. At Kalyazin there was only one
section of 4096~channels, with the bandwidth 3.125~MHz, so the measurements of the
right hand and left hand circular polarizations had to be made sequentially. Additionally
several records on the magnetic tape were made in the VLBI system
S-2 on April~12,~2001 at 64-m radio telescope in Bear Lakes (Russia) and July~22,~2004
at Kalyazin. (The bandwidth in both cases was 8~MHz). The spectral analysis was performed
with a software correlator, which alowed us to vary resolution in a wide range. The sensitivity
of the Nan\c{c}ay radio telescope was 1.0~K/Jy, and that of Bear Lakes and Kalyazin radio telescopes
was 0.65~K/Jy. The flux density calibration of Bear Lake and Kalyazin radio telescopes was performed with
continuum sources Cyg~A and 3C~274. The beam width of the Nan\c{c}ay radio telescope was
3$^{\prime}$.5 in the East-West direction and 19$^{\prime}$ in North-South direction.
Bear Lakes and Kalyazin radio telescopes had circular beams with the width 8$^{\prime}$.
In addition we have analyzed also observational data of this source
in OH lines from NRAO VLBA archive for November~2000 and January~2001
(observations by Fish and Migenes).

\section*{Results}

Spectra of W75~N obtained with the VLBA in  1998 (\citealp{slysh02}) and in
2001 (VLBA archive), at Nan\c{c}ay in 2003, and in Kalyazin in 2004, in the OH line 1665~MHz
in right hand circular polarization are presented on Fig.~1. Some differences of spectra
obtained with the VLBA from the rest three spectra  arise because they were obtained
in the VLBI regime while the single dish spectra were obtained in the total power regime.
We do not show total power spectra from individual VLBA antennae, because they are corrupted
by the contribution from nearby maser W75~S due to a large beam of VLBA antennae. In the first
spectrum of 1998 which was obtained during VLBA mapping several features were observed which
were marked as A,C,E,F,G,J~and~K in \citealp{slysh02}. They were present also in earlier
observations of 1983 (\citealp{baart}), approximately with the same intensity except features
J~and~K at radial velocities 3.0~km/s and 0.65~km/s which had variable intensity. Subsequent VLBA
observations in November 2000 and January 2001 (VLBA archive) have shown that the spectrum of the source
in the radial velocity range from 4~km/s  to 14~km/s, remained unchanged, while at lower radial
velocities significant changes took place. A strong spectral feature at the radial velocity
2.3~km/s and with a flux density 145~Jy (75~Jy  in the right hand circular polarization which
is shown in Fig.~1)  appeared, with the line width 0.35~km/s as well as a weaker feature at
the radial velocity  $-$0.2~km/s with a flux density 35~Jy. We marked them as P$_1$ and P$_2$,
respectively. At the same time the features J~and~K have diminished by a factor from 2 to 3.
Subsequent observations in April~2001 with a 64-m telescope at Bear Lakes (spectrum is not shown)
demonstrated that the flared features P$_1$ and P$_2$ dimmed by a factor from~2~to~5, remaining
at the same radio velocities. One can suppose that the flare of the features P$_1$ and P$_2$ was
a precursor of a still brighter flare of the maser emission which has been first observed by us
2.5~years later, on October~24,~2003 with the Nan\c{c}ay radio telescope. At the spectrum on Fig.~1 one
can see a very bright feature at the radial velocity 1.8~km/c and the flux density 750~Jy and line
width  0.35~km/s (N$_1$) as well as two weaker features N$_2$ at the radial velocity 0~km/s and
N$_3$ at  $-$1~km/s. Polarization measurements at Nan\c{c}ay show that all the features of the flare
N$_1$, N$_2$, N$_3$ have high degree of linear polarization exceeding ($\geq$80\%). The difference
between radial velocities of the new features N$_1$, N$_2$ and those of the precursor which amounts to about
$-$0.5~km/s, and the decrease of the precursor in April~2001 mean that a flare of new spectral features
was observed rather than the brightening of old ones. At the same time the features from~A~to~G have the
same flux density and radial velocity. The features J~and~K, even if present, have the flux density
less than 10~Jy. Subsequent observations conducted at Kalyazin in July~2004 confirmed the existence
of the flared features practically at the same radial velocities and with the same flux densities. Later
observations at Kalyazin in October-December~2004 reveal some decay of the strongest spectral feature
N$_1$ accompanied by an increase of the weaker flare features N$_2$ and N$_3$. In the second OH
line at the frequency 1667~MHz, observed at Nan\c{c}ay, new spectral features have appeared also, but
the intensity did not exceed 20~Jy and was comparable with the intensity of the rest of spectral feature.

\section*{Discussion}

Although the variability of OH masers is a well established phenomenon, there was no reports about
OH maser emission flares which such a large flux density. Moreover, at present there is no known
OH maser with a flux density equal or exceeding that of W75~N. A powerful flare of maser emission
has been  observed twice in the line of water vapor at  1.35~cm in Orion
(\citealp{abr};~ \citealp{matv};~\citealp{omodaka}) with the flux above
10~MJy. Since the flux of water masers exceeds the flux of
OH masers by a factor of 10~---~10$^3$, there is no surprise that the scale of the flare is respectively
larger for water maser.

In the paper of Slysh et al.~(2002) a model of W75~N maser was proposed, in which
the maser spots are located in disks rotating around two massive stars which in turn excite
ultracompact HII-regions VLA~1 and VLA~2 (\citealp{tor}) (Fig.~2).
Spectral features from A~to~I at the radial velocities from 12.45~km/s to 3.70~km/s are connected with
VLA~1 (some weaker features are not visible on the scale of our diagram and have no marks);
only two features are connected with VLA~2: J~and~K at the radial velocities 3.0~km/s and 0.65~km/s, respectively.
Judging on radial velocities of new spectral features of the flare 2~km/s and 0~km/s, they could be also
connected with the VLA~2. VLA~1 and  VLA~2 are separate centers of maser emission also in lines of H$_2$O and,
probably, of methanol. These ultracompact HII regions are excited by young massive stars of spectral
type B1 or B2 with a mass 10~M$_\odot$ (\citealp{shep}). The separation between the stars is about 1400~AU.
In the paper of Torrelles et al. (\citealp{tor}) a model of the source W75~N was proposed, in which the
stars are sources of a wind and a shock which excite the maser emission. Maser system connected with
VLA~2 is more compact and, apparently, more active. The evidence of this comes from observations of
high velocity masers in the OH 1667~MHz line in 1986 (\citealp{hut}) and of the variability of H$_2$O
maser emission with a period of 11.5~years, which are associated with VLA~2 (\citealp{lekht}). The  new
flare in the OH 1665~MHz line with the flux 1000~Jy, which is reported in this paper has occurred also
in the maser system connected with the VLA~2. In the framework  of the model of stellar wind shock one
can assume that the flare began upon arrival of the shock into a region of enhanced abundance of H$_2$O
 and hydroxyl molecules and elevation of the kinetic temperature in this region, or was triggered by
the enhancement of abundance of H$_2$O and hydroxyl molecules caused by the shock.

Almost simultaneously with the rising of the flux of spectral features of the precursor flare spectral
features at the radial velocity 2.3~km/s è $-$0.2~km/s, a drop of the flux from spectral features J~and~K
at the radial velocities 3.0~km/s and 0.65~km/s has occured. One can see from Fig.~1 that these features
are not visible in the spectrum of W75~N of 2001, at least at the level of 10~Jy,
while in 1998 they had flux comparable  with that of the
brightest feature~A (\citealp{slysh02}). The features J~and~K were brighter in the past: the maximum flux
of feature~J was observed in 1993 (\citealp{hut}); the feature was weaker in the earlier epochs of 1983
and 1986  (\citealp{baart};~\citealp{hut}), as well as in later epoch 1998 (\citealp{argon};~\citealp{slysh02}).
Using this data one can estimate that the duration of feature~J emission was t=11~years at the half intensity
level, with the maximum in 1993.  It is known from the VLBA maps (Fig.~2) (\citealp{slysh02}) that the
feature~J as well as feature~K is extended and has a shape of a thin strip 10.7$\times$2.4 milliarcsec in extent,
with the position angle 114$^\circ$. At the distance to W75~N 2~kpc this corresponds to the linear size of the
maser condensation l$_1$$\times$l$_2$=(21.4$\times$4.8)~AU. If one assumes that the emission from the
feature~J was caused by passing of a shock across the clump of matter with the above mentioned size, the velocity
of the shock must be not less than l$_1$/t=9.3~km/s, if the clump is in the plane of the sky, or more with
an account of the projection effect.

Normal supersonic shocks are hardly able to propagate with such a velocity, since the sound velocity
in molecular clouds is small, less than 1~km/s. However magneto hydrodynamic waves propagating with the
Alfven velocity may well be the agent exciting maser emission. It is known that in the maser emission
regions magnetic field may exceed 5$\times$10$^{-3}$~Gauss, as follows from measurements of the Zeeman
splitting. For example in W75~N two maser condensations have magnetic field 5.2$\times$10$^{-3}$~Gauss
and 7.7$\times$10$^{-3}$~Gauss, respectively (\cite{slysh02}). Assuming the value of magnetic field equal to
H=3$\times$10$^{-3}$~Gauss  and gas density n$_{H2}$=10$^{5}$--10$^{6}$~cm$^{-3}$, one gets Alfven velocity
15~km/s  and  5~km/s, respectively, in agreement with the estimates of the excitation propagation velocity
which ignited the maser emission of component~J. One can suppose that after passing through the component~J
the excitation with the Alfven velocity has arrived to the next condensation which gave start to the next
flare of the maser emission. Its intensity this time happened to be about 50~times higher than the
intensity of component~J; this can be explaned by a more favorable conditions for the maser emission.
More detailed information about temporal evolution of the maser emission flare can be obtained from results
of high angular resolution VLBI mapping and from the space interferometer "Radioastron".

\section*{Conclusion}

The brightest flare of emission in OH line  since the beginning of observations in 1965 was discovered.
The flare is connected with the activity of a young massive star which excites ultracompact  HII region
VLA~2 in the star forming region W75~N. The onset of the flare was accompanied by attenuation of emission
from other maser condensations connected with the  VLA~2. Such a behavior can be explained by propagation
of magneto hydrodynamic shock through a region of enhanced gas concentration.

\begin{acknowledgments}
Authors are grateful to collaborators from Astro Space Center of the Lebedev Physical Institute
B.Z.~Kanevskii, A.I.~Smirnov, S.V.~Logvinenko, V.I.~Vasil'kov, S.V.~Kalenskii,
 and also to Ray Escoffier (NRAO, USA) for the development of the receiver and correlator,
 and to Jean-Michel~Martin, Eric~Gerard (Meudon Observatory, France), Victor~Migenes (University of
 Guanojuato, Mexico) for the help with observations, and to data analysts of NRAO for the access
 to the archive of results of observations.

This research was conducted  under a partial support from the Russion Foundation for Basic
Research (project code N~04-02-17057), CRDF (grant N~RP1-2392-MO-02), Program for Fundamental
Research  of the Russian Academy of Science "Extended Objects
in the Universe", Federal Program "Astronomy", and also in the framework of the preparation the
observational program for the space interferometer "Radioastron".
\end{acknowledgments}

%================== LITERATURE ================================================

\clearpage
\newpage

\begin{figure}

\vskip -5mm \setcaptionmargin{5mm} \onelinecaptionstrue
\hskip-20mm\includegraphics[width=0.7\linewidth]{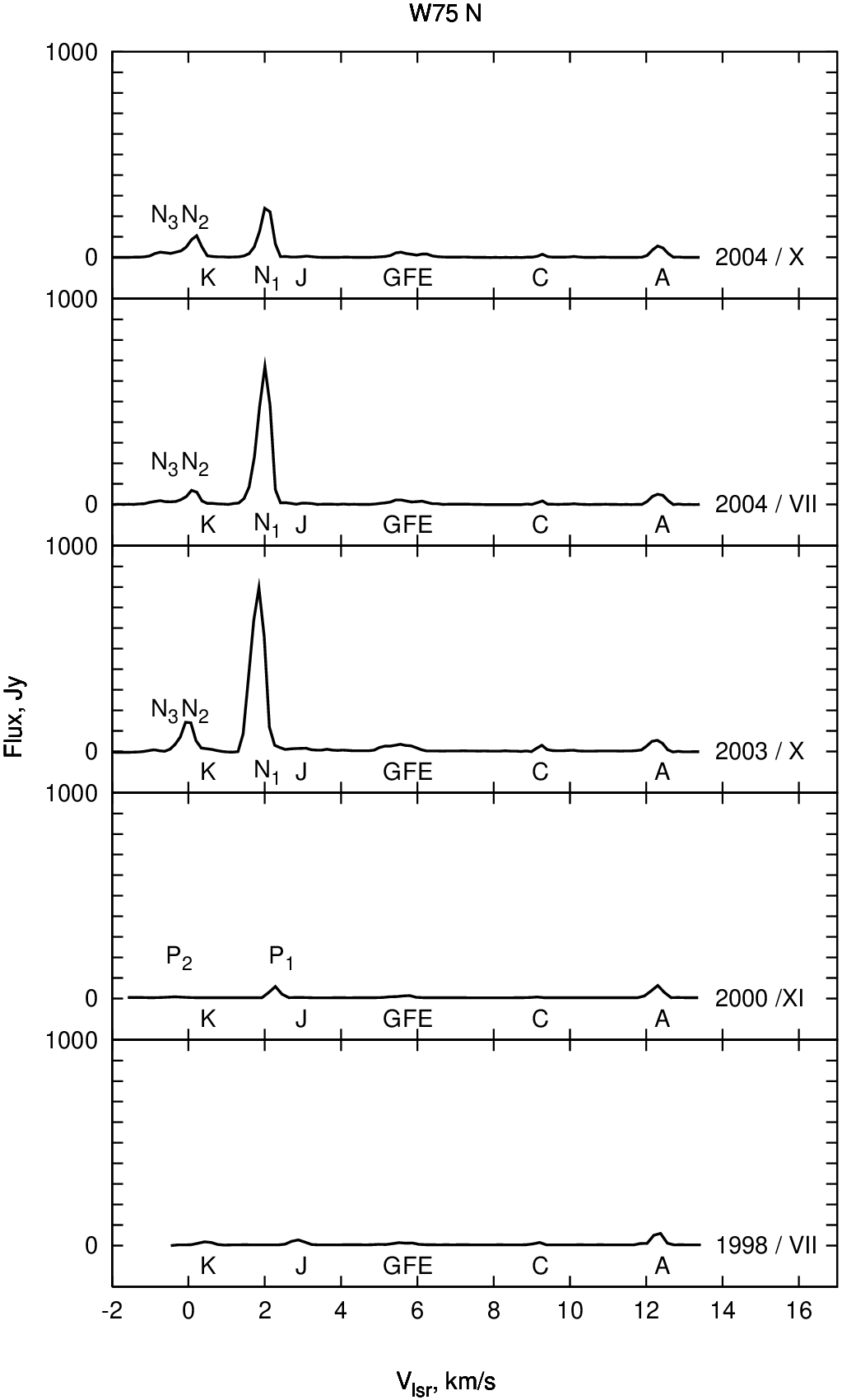}
\vskip 5mm \captionstyle{normal} \caption{Spectra of OH maser W75~N in 1665~MHz
line (right hand circular polarization), obtained at different epochs.}
1998/VII: VLBA array, July~1, 1998;
2000/XI: VLBA array, November~22, 2000;
2003/X: Nancay radio telescope, October~24, 2003;
2004/VII:  Kalyazin radio telescope, July~14, 2004;
2004/X: Kalyazin radio telescope, October~20, 2004;
Spectral features marked by letters A~---~K, were identified in the paper of Slysh et al. (2002),
features marked by letters P$_1$, P$_2$ and N$_1$,
N$_2$ N$_3$ are new spectral features of the precursor (P) and flare (N).
VLBA spectra are shown in interferometric mode, in which contribution from the nearby maser
W75~S at the distance 14$^{\prime}$ is canceled.

\end{figure}

\clearpage
\newpage

\begin{figure}

\vskip 20mm \setcaptionmargin{5mm} \onelinecaptionstrue
\hskip-20mm\includegraphics[width=0.5\linewidth]{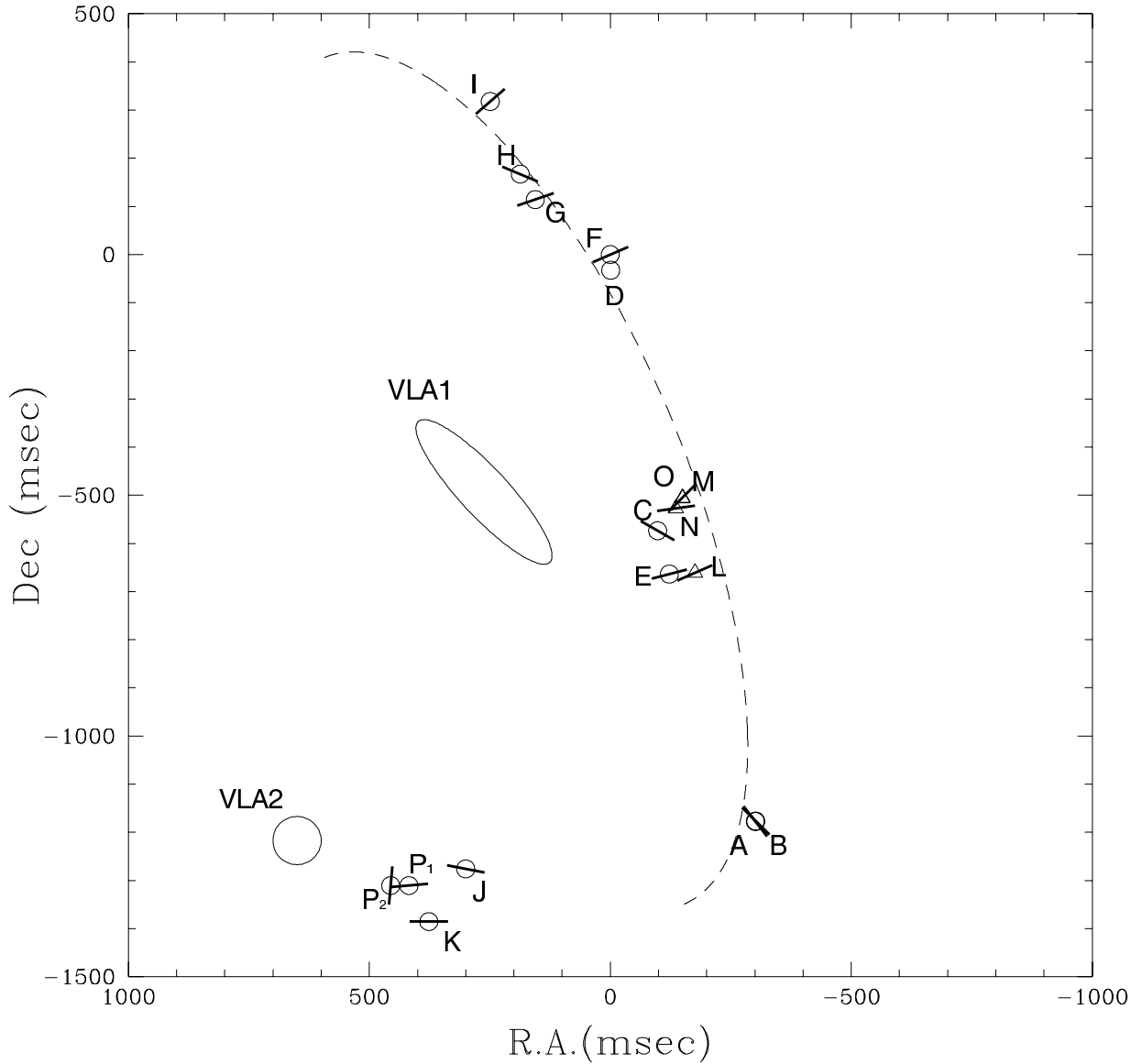}
\vskip 20mm \captionstyle{normal} \caption{
Position of precursors P$_1$ and P$_2$ (in the vicinity of VLA~2) and the direction of
the magnetic field in them (determined from VLBA archive observations in November~2000
and January~2001, Slysh, private communication).
Features P$_1$ and P$_2$ are plotted on the VLBA map of OH maser condensations in W75~N,
taken from Slysh et al. (2002).
Circles show  1665~MHz features, triangles show 1667~MHz features.
The direction of the magnetic field is shown by the vectors.
The position of the ultracompact HII regions VLA~1 and VLA~2 are shown by an ellipse and a large
circle, respectively. Dashed line shows a disk possibly connected with  VLA~1.}
\end{figure}

\end{document}